\documentclass[aps,prb,twocolumn,showpacs,preprintnumbers,amsmath,amssymb,superscriptaddress]{revtex4}

\usepackage{graphicx}
\usepackage{dcolumn}
\usepackage{bm}

\begin{document}
\preprint{PREPRINT (\today)}

\title{Probing the Yb$^{3+}$ spin relaxation in Y$_{0.98}$Yb$_{0.02}$Ba$_{2}$Cu$_{3}$O$_{x}$ by Electron Paramagnetic Resonance}
\author{A.~Maisuradze}
\affiliation{Physik-Institut der Universit\"{a}t
Z\"{u}rich, Winterthurerstrasse 190, CH-8057 Z\"{u}rich, Switzerland}
\author{A.~Shengelaya}
\affiliation{Institute of Physics, Tbilisi State University, Chavchavadze av. 3, GE-0128 Tbilisi, Georgia}
\author{B. I.~Kochelaev}
\affiliation{Department of Physics, Kazan State University, Kazan, 420008,
Russia}
\author{E.~Pomjakushina}
\affiliation{Laboratory for Developments and Methods, Paul Scherrer Institut, CH-5232 Villigen PSI, Switzerland}
\affiliation{Laboratory for Neutron Scattering, ETH Z\"{u}rich and Paul Scherrer Institut, CH-5232
Villigen PSI, Switzerland}
\author{K.~Conder}
\affiliation{Laboratory for Developments and Methods, Paul Scherrer Institut, CH-5232 Villigen PSI, Switzerland}
\author{H.~Keller}
\affiliation{Physik-Institut der Universit\"{a}t
Z\"{u}rich, Winterthurerstrasse 190, CH-8057 Z\"{u}rich, Switzerland}
\author{K.A.~M\"uller}
\affiliation{Physik-Institut der Universit\"{a}t Z\"{u}rich,
Winterthurerstrasse 190, CH-8057 Z\"{u}rich, Switzerland}
%

\begin{abstract}

The relaxation of Yb$^{3+}$  in YBa$_{2}$Cu$_{3}$O$_{x}$ ($6<x<7$) was studied using Electron Paramagnetic Resonance (EPR). It was found that both electronic and phononic processes contribute to the Yb$^{3+}$ relaxation. The phononic part of the relaxation has an exponential temperature dependence, which can be explained by a Raman process via the coupling to high-energy ($\sim$500 K) optical phonons or an Orbach-like process via the excited vibronic levels of the Cu$^{2+}$ ions (localized Slonczewski-modes). In a sample with a maximum oxygen doping $x$=6.98, the electronic part of the relaxation follows a Korringa law in the normal state and strongly decreases below $T_{c}$. Comparison of the samples with and without Zn doping proved that the superconducting gap opening is responsible for the sharp decrease of Yb$^{3+}$ relaxation in  YBa$_{2}$Cu$_{3}$O$_{6.98}$. It was shown that the electronic part of the Yb$^{3+}$ relaxation in the superconducting state follows the same temperature dependence as $^{63}$Cu and $^{17}$O nuclear relaxations despite the huge difference between the corresponding electronic and nuclear relaxation rates.

\end{abstract}

\maketitle
\section{INTRODUCTION}
It is well known that in the  high-$T_{c}$ superconductor YBa$_{2}$Cu$_{3}$O$_{x}$ the substitution of yttrium by  isovalent rare-earth (RE) ions, having local magnetic 4$f$ moments, does not change the critical temperature $T_{c}$ considerably.\cite{Hor} This makes it possible to use these local moments as useful paramagnetic probes of the electronic states within the CuO$_{2}$ planes without seriously perturbing them.
Relaxation of the RE magnetic moments  provides important information about fluctuating electric and magnetic fields in cuprate superconductors. In the normal state, RE ions with non-zero orbital moment can interact with phonons, spin fluctuations and charge carriers. These interactions limit the life-time of the crystal field (CF) excitations and lead to broadening of the observed CF transitions. Inelastic neutron scattering (INS) is widely used to study the relaxation of the RE magnetic moments in cuprates superconductors by measuring the linewidths of the observed CF transitions.\cite{Mesot} However, the mechanism of the relaxation of RE ions in cuprates is the issue of hot debates.\cite{Boothroyd1,Staub1}
Most of the authors consider an interaction of 4$f$ spins with charge carrier spins (Korringa mechanism) as a dominant channel of relaxation.\cite{Walter,Amoretti,Boothroyd2,Furrer}  Other authors, in opposite, conclude that the interactions of RE spins with charge carriers are negligible and that the interactions with lattice vibrations are only responsible for the relaxation behavior of the 4$f$ spins in cuprates.\cite{Staub2,Staub3,Roepke}  At the moment there is no consensus on this subject and therefore it is important to disentangle the electronic and phononic sources of relaxation.

Relaxation of the 4$f$ magnetic moments can be studied also by measuring the linewidth of the Electron Paramagnetic Resonance (EPR) signal of the RE ions.\cite{Barnes} There have been a few EPR studies of the RE ions with non-zero orbital moment (Er and Yb) doped in cuprates.\cite{Elschner,Eremin,Kurkin,Shimizu1,Shimizu2,Ivanshin1,Ivanshin2,Gafurov}  Again, the situation with EPR is similarly contradictory as in INS studies. Some groups observed both electronic  (Korringa) and phononic (Orbach) contributions to EPR relaxation.\cite{Elschner,Shimizu1,Shimizu2}  While others conclude that the Korringa contribution is negligible and the interaction with lattice vibrations only is sufficient to explain the temperature dependence of the EPR linewidth in cuprates.\cite{Ivanshin1,Ivanshin2,Gafurov}

In this work we report a detailed EPR study of the Yb$^{3+}$ relaxation in YBa$_{2}$Cu$_{3}$O$_{x}$ samples with oxygen content $x$ covering the whole doping range ($6<x<7$). By measuring the temperature dependence of the Yb$^{3+}$ EPR linewidth in a broad temperature range, it was found that {\itshape both} phononic and electronic mechanisms contribute to the relaxation. The electronic contribution decreases with decreasing oxygen content $x$, while the phononic contribution is practically doping independent and has an exponential temperature dependence. In a sample with a maximum oxygen doping $x$=6.98, the electronic part of the relaxation follows a Korringa law in the normal state, and a sharp drop of the relaxation rate is observed below $T_{c}$. In the superconducting state the electronic part of the Yb$^{3+}$ relaxation rate follows the same temperature dependence as $^{63}$Cu and $^{17}$O nuclear relaxations.

This paper is organized as follows: Sample preparation and experimental details are described in Sec. II. The EPR spectra and the procedure of their analysis are discussed in Sec.  III. Sec. IV presents the temperature  and doping dependence of the Yb$^{3+}$ relaxation rate in the normal and superconducting states. In Sec. V we summarize our results and the  conclusions of this study.

\section{Experimental details}

The polycrystalline samples of Y$_{1-y}$Yb$_{y}$Ba$_{2}$Cu$_{3}$O$_{x}$
were prepared by the standard solid state reaction by using Y$_{2}$O$_{3}$, Yb$_{2}$O$_{3}$, BaCO$_{3}$ and CuO of a minimum purity of 99.99\%. Appropriate amounts of starting reagents were mixed and calcinated at temperatures 800-920$^{o}$C during at least 150 h in air, with several intermediate grindings. Finally, the as-prepared sample was oxidized in oxygen atmosphere (1 bar of O$_{2}$) at 500$^{o}$C. After the oxidation the sample had an oxygen content close to 7 (6.98). The required oxygen content in the samples was adjusted by gettering in a closed ampoule with metallic copper (850$^{o}$C, 10 h; cooling 10$^{o}$C/h). The oxygen content in the reduced samples was checked by comparing theoretical and real mass changes of the oxidized getter and the reduced sample. For all the samples phase purity was checked with a conventional X-ray diffractometer (SIEMENS D500).

The dilute level of the Yb doping ($y$=0.02) was chosen in order to minimize broadening effects from Yb-Yb interactions and at the same time to obtain a sufficiently strong EPR signal. The EPR measurements were performed with an $X$-band BRUKER EMX spectrometer equipped with an Oxford Instruments helium flow cryostat.  In order to avoid a signal distortion due to skin effects, the samples were ground and the powder was suspended in epoxy. The $c$-axes grain-orientation was obtained by placing the samples in a 9 T magnetic field until the epoxy hardened. As a result of the orientation procedure the $c$-axes of the grains were preferentially aligned along the magnetic field direction.\cite{Farrell}

EPR spectra were measured in five samples Y$_{0.98}$Yb$_{0.02}$Ba$_{2}$Cu$_{3}$O$_{x}$ with $x$=6.1, 6.4, 6.5, 6.6, 6.98 and critical temperatures $T_{c}$ of 0, 12(1), 51(1), 60(1), and 93(1) K, respectively. In addition, one Zn-doped sample  Y$_{0.98}$Yb$_{0.02}$Ba$_{2}$(Cu$_{0.97}$Zn$_{0.03}$)$_{3}$O$_{6.95}$ was measured with $T_{c}$=57(1) K. 

\begin{figure}[htb]
\includegraphics[width=0.70\linewidth]{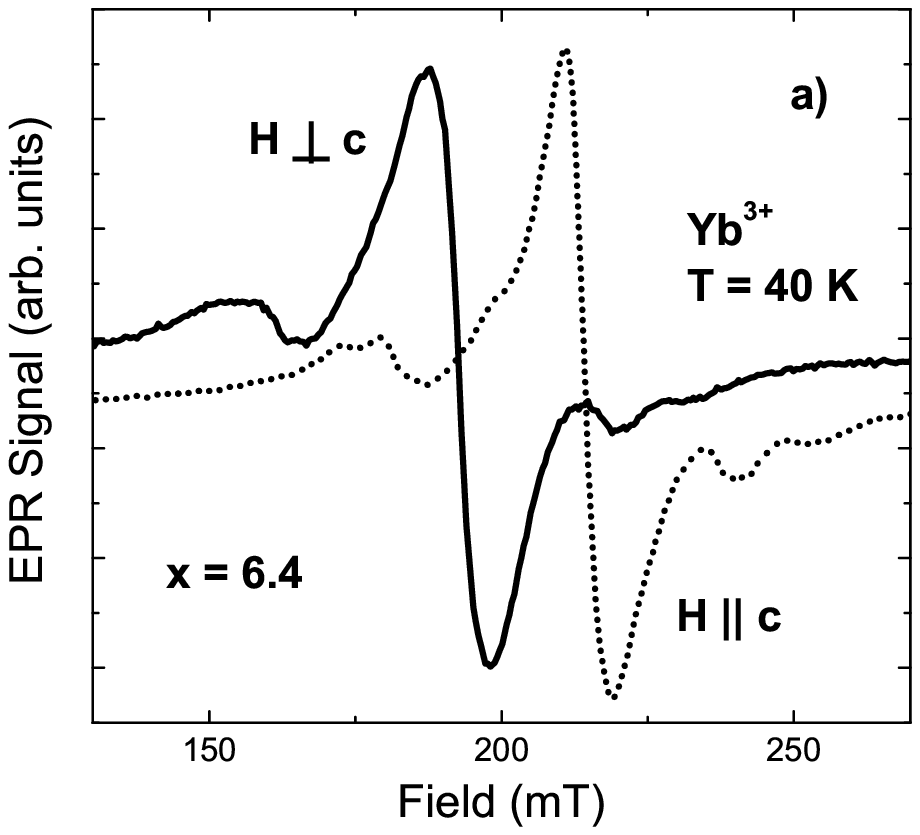}
\includegraphics[width=0.70\linewidth]{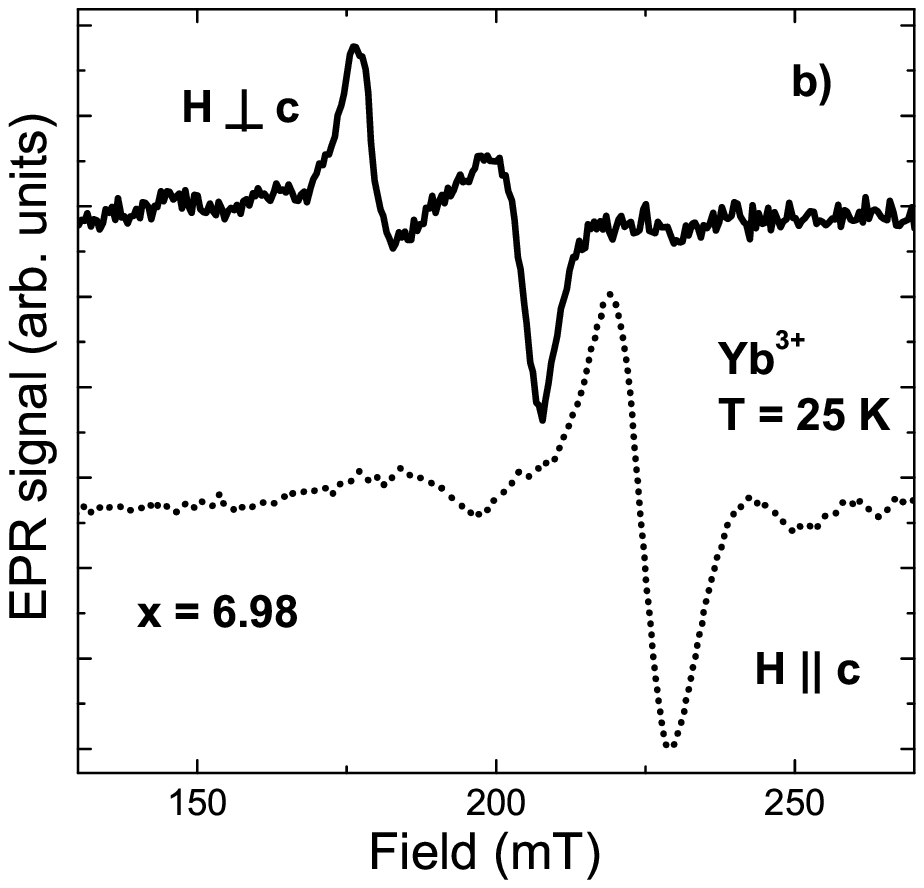}
\caption{EPR spectra of Yb$^{3+}$ in grain oriented Y$_{0.98}$Yb$_{0.02}$Ba$_{2}$Cu$_{3}$O$_{x}$  with different oxygen contents: (a) $x$ = 6.4; (b) $x$ = 6.98. Two orientations correspond to the external magnetic field along and perpendicular to the crystal $c$-axis.}\label{Fig.1}
\end{figure}

\section{Analysis of the ytterbium EPR spectra}

In the YBa$_{2}$Cu$_{3}$O$_{x}$ structure, the rare-earth site is eight-fold coordinated by oxygens lying in the CuO$_{2}$ bilayers. Group theoretical considerations show that the eight-fold degeneracy of the ground-state multiplet $^{2}$F$_{7/2}$ of the Yb$^{3+}$ ions (4$f^{13}$) is split by the crystal electric field of orthorhombic symmetry into four Kramers doublets.\cite{Bleaney}  Generally, this splitting is large enough so that only the lowest lying doublet is appreciably populated at low temperatures. In fact, inelastic neutron scattering measurements showed that in YbBa$_{2}$Cu$_{3}$O$_{7}$ the first excited doublet lies 1000 K above the ground doublet.\cite{Guillaume} So, EPR signals would be observed only for this doublet, which can be described in an effective $S$ = 1/2 notation.

Fig. 1(a) shows Yb$^{3+}$ EPR spectra of Y$_{0.98}$Yb$_{0.02}$Ba$_{2}$Cu$_{3}$O$_{6.4}$ at $T$=40 K with the external magnetic field along and perpendicular to the crystal $c$-axis. The spectra correspond to Yb$^{3+}$ with an effective spin $S$=1/2 with $g$ values $g_{\parallel}$=3.13(3) and $g_{\perp}$=3.49(3). The average value  $g_{av}$=3.37 is close to $g$=3.43 expected for the isolated $\Gamma_{7}$ ground doublet.\cite{Bleaney} Fig.1(b) shows Yb$^{3+}$ EPR spectra for the optimally doped sample $x$=6.98 at $T$=25 K. Note that the EPR line splits into two components for the $H\bot c$ orientation. This reflects the transition from tetragonal to orthorhombic crystal symmetry in YBa$_{2}$Cu$_{3}$O$_{x}$ with increasing oxygen content. Enhanced noise seen in these spectra is related to the superconducting state. It is known that in the superconducting state the strong noise is generated due to vortex motion in the modulating magnetic field used in standard EPR spectrometers.

Fig. 2 shows typical EPR spectra for the $x$=6.4 sample at different temperatures with the magnetic field applied perpendicular to the $c$-axis. The EPR lines in the $g\approx2$ region are due to Cu$^{2+}$ defect centers, which are always present in YBa$_{2}$Cu$_{3}$O$_{6+x}$.\cite{Kochelaev}
The concentration of the Cu$^{2+}$ defect centers in our samples correspond to 1-2\% of the total copper content, as determined from the Cu$^{2+}$ EPR signal intensity.
The Yb$^{3+}$ EPR spectrum at 40 K is rather complex, with a dominant central line and shoulders on each side. The shoulders are due to partially resolved hyperfine components. Natural ytterbium has 69\% even-mass isotopes with nuclear spin $I$=0, 14\% $^{171}$Yb with $I$=1/2, and 16\% $^{173}$Yb with $I$=5/2. Both odd isotopes give rise to nuclear hyperfine interactions.
In Fig. 2 one can see that with increasing temperature the Yb$^{3+}$ EPR line broadens and the multiple-line structure of the Yb spectra gradually merges into one line which continues to broaden with temperature. This broadening is due to the Yb relaxation and is the subject of the present study.

\begin{figure}[htb]
\includegraphics[width=0.8\linewidth]{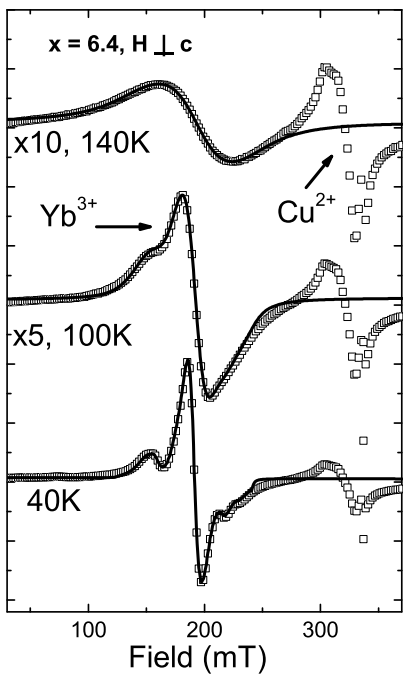}
\caption{EPR spectra in grain oriented Y$_{0.98}$Yb$_{0.02}$Ba$_{2}$Cu$_{3}$O$_{6.4}$ at different temperatures for a magnetic field direction perpendicular to the crystal $c$-axis. The solid lines are fits to the data, including relaxation as described in the text.
}\label{Fig.2}
\end{figure}

The observed line shape can be interpreted in the following way: a temperature-independent residual function due to inhomogeneous broadening which is further broadened by a temperature dependent relaxation. The method by which the relaxation-induced peak-to-peak derivative linewidth $\Delta B_{pp}^{r}$ is extracted from the total peak-to-peak linewidth of the EPR signal $\Delta B_{pp}^{t}$ depends upon the line shapes associated with inhomogeneous and homogeneous broadening mechanisms. The homogeneous broadening leads to Lorentzian line shape while the inhomogeneous low-temperature line shape is usually approximated as either Lorentzian or Gaussian. For a Lorentzian-Lorentzian convolution there is a simple relation : $\Delta B_{pp}^{r}(T) = \Delta B_{pp}^{t}(T) - \Delta B_{pp}^{0}$, where $\Delta B_{pp}^{0}$ is a residual, temperature-independent linewidth. The result of a Lorentzian-Gaussian convolution is called a Voigt function, which cannot be expressed in closed form. 

In our case the low-temperature line shape is neither Lorentzian nor Gaussian. The  anisotropic nature of the Yb$^{3+}$ signal, the partially resolved hyperfine structure, and the non-ideal grain alignment, make it very difficult to accurately model the complex shape of the Yb EPR spectra and its evolution with temperature. Therefore, in order to extract the linewidth related to relaxation, we used an approach similar to one used in inelastic neutron scattering studies of rare-earth relaxation in cuprates.\cite{Boothroyd2} 

The inhomogeneous broadening of an EPR line is due to the spread of the resonance frequencies of an assembly of electronic spins. The width of the single spin packet contributes to homogeneous broadening. This situation can be described by a convolution integral,\cite{Smirnov} 
\begin{equation}
 I(B) = \int p(B')f(B-B')dB'=p(B)*f(B)
 \end{equation} 
where * is the convolution symbol, $p(B)$ is the inhomogeneous line shape and $f(B)$ is the resonance line profile of the spin packet given by a Lorentzian line function. 
EPR spectra are usually detected in the form of a first derivative, $I'(B)=\partial I(B)/\partial B$. In this case $I'(B)=p(B)*f'(B)$, where $f'(B)$ is the first derivative of a Lorentzian with amplitude $A$, center at $B_{0}$ and the peak-to-peak width $\Delta B_{pp}^{r}$ related with relaxation,

\begin{equation}
 f'(B) = A\frac{2\cdot(B-B_0)}{\Delta B_{pp}^{r}} \cdot \left[3+\left(\frac{2\cdot(B-B_0)}{ \Delta B_{pp}^{r}}\right)^2\right]^{-2}.
\end{equation} 

In order to describe the inhomogeneous line shape $p(B)$, we divided Yb EPR spectra in 120 points with a step of $\delta B$=1 mT in the interval of 120-240 mT. In this case an integral is replaced by a sum of 120 Lorentzians:
\begin{equation}
 I'(B) = \sum_{i=1}^{120} p_i(B_i) \cdot f'(B-B_i)\delta B
 \end{equation} 
The residual function $p_{i}(B_{i})$, characterizing the inhomogeneous line shape was obtained at low temperatures where relaxational broadening is negligible. We found that in all samples measured in the present work, except one Zn-doped sample, the relaxational broadening was absent below 40 K. Therefore, this
temperature was used to determine the inhomogeneous line shape for samples without Zn doping. In Zn-doped sample the linewidth continues to decrease down to 15 K and therefore the inhomogeneous line shape was determined at this temperature. As an example, the line drawn through the 40 K data in Fig. 2 represents the residual function for $x$=6.4 sample. Having established the residual function, we kept the coefficients $p_{i}$ fixed at all temperatures and fitted the data at high temperatures by convolving the residual function with the broadening function of Lorentzian shape. The center $B_{0}$, width $\Delta B_{pp}^{r}$, and amplitude $A$ of the broadening function were the only variable parameters. The resulting fits are shown in Fig. 2.

The peak-to-peak linewidth $\Delta B_{pp}^{r}$ is related to the transverse relaxation rate $T_{2}^{-1}$ as follows:\cite{Stapleton} 
\begin{equation}
T_{2}^{-1} = \frac{\sqrt{3}g\mu_{B}\Delta B_{pp}^{r}}{2\hbar}=7.62\times 10^{6}g\Delta B_{pp}^{r}                 
\end{equation}
For paramagnetic relaxation of RE ions  in cuprates $T_{2}^{-1}$ is equal to the spin-lattice relaxation rate $T_{1}^{-1}$, as was shown for Gd$^{3+}$ in 
Y$_{0.99}$Gd$_{0.01}$Ba$_{2}$Cu$_{4}$O$_{8}$.\cite{Atsarkin1}

\section{Temperature and doping dependence of the ytterbium relaxation}

\subsection{Yb$^{3+}$ relaxation in the normal state}

Figure 3 shows the temperature dependence of the peak-to-peak width $\Delta B_{pp}^{r}$ related with relaxation and corresponding relaxation rate $1/T_{1}$ of Yb$^{3+}$ in Y$_{0.98}$Yb$_{0.02}$Ba$_{2}$Cu$_{3}$O$_{x}$ for different oxygen contents. In order to distinguish between electronic and phononic contributions to relaxation, let us first consider  the sample with the lowest oxygen content ($x$=6.1). One can see in Fig. 3 that  in this sample, which is antiferromagnetic and insulating with practically no charge carriers, the relaxation rate is comparable to those of samples with much higher oxygen contents. Consequently, this suggests that the phonon contribution to the rare-earth relaxation rate is significant at all oxygen doping levels, since in the x=6.1 sample the electronic contribution is expected to be negligible.

\begin{figure}[htb]
\includegraphics[width=1.0\linewidth]{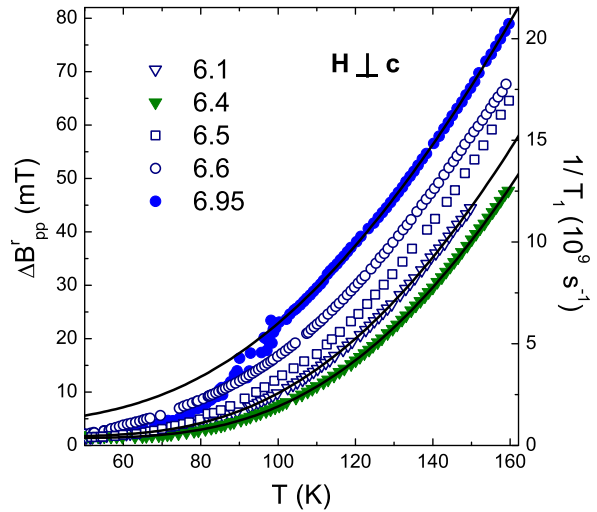}
\caption{(Color online). Temperature dependence of the width of the broadening function
$\Delta B_{pp}^{r}$ and the corresponding spin-lattice relaxation rate $T_{1}^{-1}$ of Yb$^{3+}$ in Y$_{0.98}$Yb$_{0.02}$Ba$_{2}$Cu$_{3}$O$_{x}$ with different oxygen content $x$ for $H\bot c$. The solid lines represent the best fit to Eq. (6).
}\label{Fig.3}
\end{figure}

We observed that the temperature dependence of the relaxation rate follows closely the exponential function $C\exp(-\Delta/T)$  with $\Delta$=520(30)~K and 570(30)~K, for $x$=6.1 and 6.4 respectively. This is demonstrated in Fig.~4, where the relaxation rates as a function of inverse temperature are plotted on a semi-logarithmic scale. Such an exponential dependence is expected for the Orbach relaxation process via an excited intermediate energy level.\cite{Orbach} In this case $\Delta$ corresponds to the separation between the ground state doublet and the excited level. According to inelastic neutron scattering experiments the first excited energy level of Yb in YBa$_{2}$Cu$_{3}$O$_{7}$ is about 1000 K above the ground state doublet.\cite{Guillaume} Since there is no excited crystal field energy level with $\Delta \sim$ 500 K, the traditional Orbach relaxation mechanism can be excluded.

\begin{figure}[htb]
\includegraphics[width=1.0\linewidth]{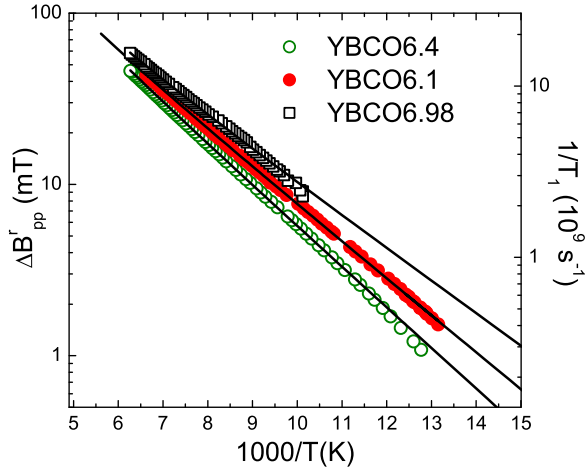}
\caption{(Color online). The width of the broadening function
$\Delta B_{pp}^{r}$ and the corresponding spin-lattice relaxation rate $T_{1}^{-1}$ of Yb$^{3+}$ in Y$_{0.98}$Yb$_{0.02}$Ba$_{2}$Cu$_{3}$O$_{x}$  
with $x$=6.1, 6.4, and 6.98 versus inverse temperature plotted on a semi-logarithmic scale. For $x$=6.98 sample only phononic contribution to relaxation is plotted. The solid lines represent the best fit to Eq. (5).}\label{Fig.4}
\end{figure}

An exponential temperature dependence of the relaxation rate is also expected for the Raman two-phonon process involving optical phonons or local vibrations.\cite{Kochelaev1,Huang} In this case the relaxation rate is
\begin{equation}
1/T_{1}=C\exp(\Omega/T)/[\exp(\Omega/T)-1]^{2},
\end{equation}
where $\Omega$ is the optical phonon frequency.\cite{Kochelaev1,Huang} The solid lines in Fig. 4 represent a best fit to the data using Eq.(5). It is obvious that the Raman process involving optical phonons can explain the phonon contribution to the Yb$^{3+}$ spin-lattice relaxation. Optical phonons within the energy range 500(50) K  exist  in YBa$_{2}$Cu$_{3}$O$_{x}$. These are: (i) the in-plane bond-bending (500-560 K) and (ii) the out of plane $B_{1g}$ (470 K) phonons.\cite{Pintschovius} Theoretical calculations are necessary in order to determine which of the two optical phonon modes (in-plane bond-bending or out of plane $B_{1g}$) mostly contributes to Yb$^{3+}$ spin-lattice relaxation.

It is interesting to note that an exponential temperature dependence of the
spin relaxation rate was observed in a previous EPR study of crystals containing Jahn-Teller (JT) transition metal ions.\cite{Muller1,Muller2} In this case relaxation takes place due to an Orbach-like process via the excited vibronic levels of the JT ion (localized Slonczewski-modes).\cite{Muller3} In our case such a scenario is also possible if Yb$^{3+}$ spin relaxation occurs due to coupling to the vibrations of surrounding CuO$_{6}$ complexes, since Cu$^{2+}$ is a strong JT ion. There are no reports on JT splitting of Cu$^{2+}$ in SrTiO$_{3}$ or similar perovskites, but in MgO and CaO it is approximately 1500 K, \cite{Muller2} which is much larger than 500 K found in the present work. However, it was observed that for Ni$^{3+}$ the JT splitting in SrTiO$_{3}$ is 2-4 times reduced compared to MgO, CaO or Al$_{2}$O$_{3}$.\cite{Muller2} A similar reduction of the JT splitting can be expected also for Cu$^{2+}$ in perovskites. In addition, in cuprates the Cu$^{2+}$ ions are situated next to each other. In such a cooperative situation the energy of the Slonczewski mode will be lower compared to the isolated Cu$^{2+}$ ions. It would be interesting to search for localized vibronic modes in cuprates using inelastic neutron scattering.

For an oxygen content $x>6.4$,  Eq.(5) cannot describe the relaxation data well, and it was necessary to take into account the Korringa relaxation mechanism where localized Yb$^{3+}$ moments couple to mobile charge carriers in the CuO$_{2}$ planes through an exchange interaction. In normal metals the Korringa relaxation has a linear temperature dependence $bT$.\cite{Barnes} The parameter $b$ is proportional to the product $\left[J_{sf}N(E_{F})\right]^{2}$, where $J_{sf}$ is the exchange integral between Yb$^{3+}$ moments and holes, and $N(E_{F})$ is the density of states at the Fermi energy. In underdoped cuprates the density of states at the Fermi level is temperature dependent due to the pseudogap opening. This leads to the nonlinear temperature dependence of the relaxation rate as demonstrated by $^{89}$Y NMR\cite{Alloul} and Gd EPR\cite{Janossy} experiments in YBa$_{2}$Cu$_{3}$O$_{x}$. No exact formula exists to describe this nonlinear temperature dependence.  Therefore we did not fit the data for  underdoped samples ($6.4<x<6.98$). However, the use of the Korringa law is justified in our optimally doped sample $x=6.98$, where the pseudogap is absent.\cite{Alloul}
The solid lines in Fig. 3 correspond to fits of the data using a sum of phononic and electronic contributions:
\begin{equation}
1/T_{1}=C\exp(\Omega/T)/[\exp(\Omega/T)-1]^{2}+bT
\end{equation}

\begin{table}[htb]

\caption[~]{The fitting parameters of the Yb$^{3+}$ relaxation 
in Y$_{0.98}$Yb$_{0.02}$Ba$_{2}$Cu$_{3}$O$_{x}$ using Eq. (6).}

\begin{center}
\begin{tabular}{lccccc}
\hline
\hline
$x$ & $T_{c}$(K) & $C$(G) & $\Omega$(K) & $b$(G/K)
\\

\hline
6.1 & - & 10800(500) & 500(30) & 0  \\

6.4 & 12(1) & 13190(500) & 540(30) & 0  \\

6.98 & 93(1) & 9090(400) & 460(30) & 1.27(10)  \\

6.95(Zn) & 57(1) & 10160(400) & 490(30) & 1.27(10)  \\
\hline
\hline
\end{tabular}
\end{center}

\end{table}

The obtained parameters  $C$, $\Omega$, and $b$ are summarized in Table~I.
It is interesting to compare the value of the Korringa constant $b$=1.27(10) G/K for $x=6.98$ with the values obtained from EPR measurements of Gd-doped YBa$_{2}$Cu$_{3}$O$_{x}$. In contrast to Yb$^{3+}$, the Gd$^{3+}$ ion has zero orbital moment  ($L$=0) and interacts very weakly  with lattice vibrations. Therefore, the Korringa relaxation due to interaction with charge carriers is the dominant process and the Korringa constant can be obtained directly from the temperature dependence of the EPR linewidth. Indeed, EPR measurements on a Gd-doped EuBa$_{2}$Cu$_{3}$O$_{6.85}$ single crystal\cite{Shaltiel} showed a linear broadening of the EPR line in the temperature range 90-300 K with the Korringa constant $b$=0.5 G/K. This value of $b$ is comparable, but smaller than our value of $b$. This is expected due to the smaller oxygen content ($x=6.85$) and consequently smaller density of states at the Fermi level $N(E_{F})$ compared to our sample ($x=6.98$). 

The comparable values of Korringa constants $b$ for Gd$^{3+}$ and Yb$^{3+}$ in YBCO shows that they have similar exchange coupling $J_{sf}$ with charge carriers. The observed Korringa constants of RE ions in YBCO are at least one order of magnitude smaller than the corresponding quantities found for RE ions in conventional metals.\cite{Barnes} The small value of $b$ for RE ions in YBa$_{2}$Cu$_{3}$O$_{x}$ is due to the weak coupling $J_{sf}$ between the RE moments and the holes in the CuO$_{2}$ planes. This explains naturally the small effect of RE magnetic moments on $T_{c}$ in YBa$_{2}$Cu$_{3}$O$_{x}$.\cite{Garifullin}

\subsection{Yb$^{3+}$ relaxation in the superconducting state}

Generally, it is difficult to measure EPR in the superconducting state because of the strong  microwave absorption and the noise due to vortex motion in the modulating magnetic field used in standard EPR spectrometers. Nevertheless, in grain-aligned samples with small grain size it was possible to observe an Yb$^{3+}$ EPR signal below $T_{c}$ (see Fig. 1(b)). This allowed us to study the temperature dependence of the Yb$^{3+}$ relaxation in the superconducting state. We observed a strong reduction of the Yb$^{3+}$ relaxation rate below $T_{c}$. This is clearly seen in Fig. 3 for $x=6.98$, where the relaxation rate falls below the theoretical line given by Eq. (6). It is natural to attribute the drop of the relaxation to the opening of the superconducting gap.

\begin{figure}[htb]
\includegraphics[width=0.7\linewidth]{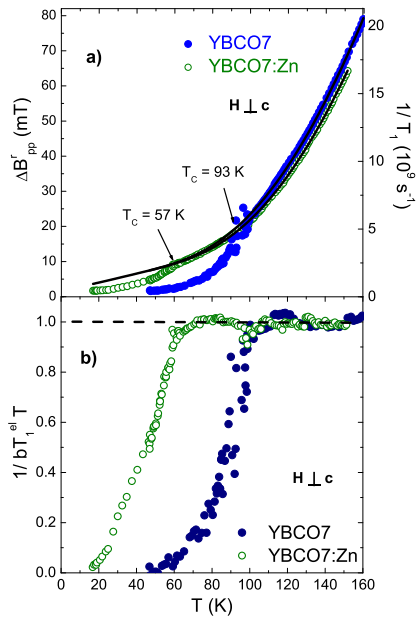}
\caption{(Color online). (a) Temperature dependence of the width of the broadening function
$\Delta B_{pp}^{r}$ and the corresponding spin-lattice relaxation rate $T_{1}^{-1}$ of Yb$^{3+}$ in Y$_{0.98}$Yb$_{0.02}$Ba$_{2}$Cu$_{3}$O$_{6.98}$ (YBCO7) and Y$_{0.98}$Yb$_{0.02}$Ba$_{2}$(Cu$_{0.97}$Zn$_{0.03}$)$_{3}$O$_{6.95}$ (YBCO7:Zn) for $H\bot c$. The solid lines represent the best fit to Eq. (6) of the YBCO7 and YBCO7:Zn data above their superconducting transition temperatures $T_{c}$=93 K and $T_{c}$=57 K, respectively. 
(b) Temperature dependence of the electronic part of the Yb$^{3+}$ relaxation $1/bT_{1}^{el}T$ in YBCO7 and YBCO7:Zn. The dashed line represents the normal state relaxation expected by the Korringa law.}\label{Fig.5}
\end{figure}

\begin{figure}[htb]
\includegraphics[width=1.0\linewidth]{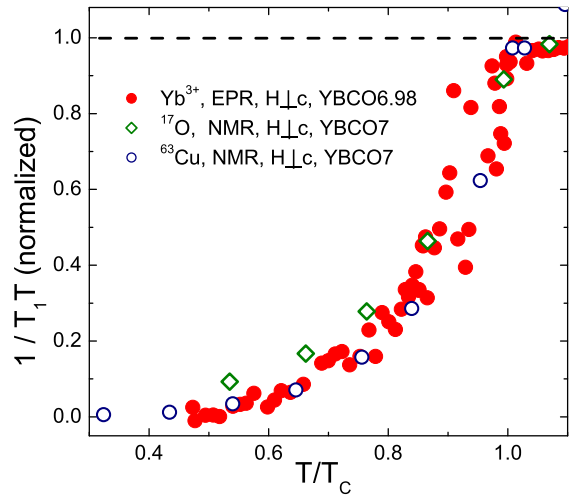}
\caption{(Color online). $(1/T_{1}T)$ of Yb$^{3+}$ in the superconducting state normalized by its value at 100 K   in Y$_{0.98}$Yb$_{0.02}$Ba$_{2}$Cu$_{3}$O$_{6.98}$  versus reduced temperature $T/T_{c}$, compared with those of $^{17}$O NMR (Ref.~\onlinecite{O-NMR})  and $^{63}$Cu NMR (Ref.~\onlinecite{Cu-NMR}) in YBa$_{2}$Cu$_{3}$O$_{7-\delta}$.}\label{Fig.6}
\end{figure}

In order to check this possibility, we measured the relaxation of Yb$^{3+}$ in Y$_{0.98}$Yb$_{0.02}$Ba$_{2}$(Cu$_{0.97}$Zn$_{0.03}$)$_{3}$O$_{6.95}$ where Zn doping reduces $T_{c}$ to 57 K. Fig.~5(a) shows the temperature dependence of the peak-to-peak width $\Delta B_{pp}^{r}$ related with relaxation and corresponding relaxation rate $1/T_{1}$ in samples with and without Zn doping. It is expected that 3\% Zn doping should not change strongly the phonon spectra and the electronic density of states. In fact, above $\sim$90 K the relaxation rates are very close for both samples. However, below $\sim$90 K the behavior of relaxation is different. While the relaxation rate of the sample without Zn doping sharply decreases below this temperature due to the onset of superconductivity, in the Zn-doped sample relaxation continues to decrease gradually until $T_{c}^{Zn}$=57 K, where a similar sharp turn is observed. This result unambiguously shows that the superconducting gap opening is responsible for the drop of Yb$^{3+}$ relaxation in  YBa$_{2}$Cu$_{3}$O$_{6.98}$. Moreover, it proves the presence of the electronic channel of relaxation described by the Korringa term $bT$ in Eq. (6). Zn doping helps to reveal the Korringa term, which is masked at high temperatures by phonon relaxation due to the much stronger temperature dependence, and below $T_{c}$ by opening of the  superconducting gap.
Figure 5(b) shows the electronic part of the Yb$^{3+}$ relaxation after subtracting the phonon contribution in pure and Zn-doped samples. The linear temperature dependence in the normal state and a sharp decrease below $T_{c}$ is clearly seen.

In superconducting samples at low temperatures extra broadening could be present due to the distribution of the diamagnetic shifts related to the irregular shape of the powder grains. However, we observed no additional broadening even in optimally doped superconducting sample down to 40 K. In fact, the superconducting and nonsuperconducting samples had the same inhomogeneous linewidth at this temperature. This is in agreement with the estimations of the diamagnetic shift $\sim$ 3 mT at 10 K in H$\perp$c orientation in YBa$_{2}$Cu$_{3}$O$_{7}$, which further decreases with increasing temperature.\cite{Janossy} Therefore, extra broadening from the distribution of this diamagnetic shift in our optimally doped sample at 40 K is estimated to be about 1 mT, which is much smaller than the observed inhomogeneous linewidth $\Delta B_{pp}$=11 mT. 

It is expected that the electronic part of the relaxation of the 4$f$ magnetic moments on the yttrium site in YBCO measured by EPR has the same temperature dependence as $^{89}$Y nuclear relaxation, since in both cases the relaxation is proportional to the imaginary part of the dynamic spin susceptibility.\cite{Kochelaev-Review} This was confirmed by EPR measurements of  Gd$^{3+}$ spin relaxation in  YBa$_{2}$Cu$_{3}$O$_{x}$ in normal state.\cite{Janossy,Atsarkin} Also, in the present work, the electronic part of the Yb$^{3+}$ spin relaxation in optimally doped YBCO above $T_{c}$ shows a linear temperature dependence (Korringa behavior) like $^{89}$Y nuclear relaxation.\cite{Alloul} Note, however, that there is a huge difference between the magnitudes (more than a factor 10$^{9}$) of the electron and nuclear relaxation rates due to the large difference between the corresponding coupling constants.

It would be interesting to compare the temperature dependences of the electron and nuclear spin relaxation rates on the yttrium site in the superconducting state, where relaxation drops due to the opening of the superconducting gap. Unfortunately, no detailed $^{89}$Y nuclear relaxation data exists for YBa$_{2}$Cu$_{3}$O$_{x}$ in the superconducting state. A very small coupling between the  yttrium nuclei and the charge carriers leads to very long relaxation times and makes $^{89}$Y NMR measurements in the superconducting state extremely difficult. Therefore, we plot in Fig. 6 the electronic part of $1/T_{1}T$ versus reduced temperature $T/T_{c}$ for Yb$^{3+}$ in $x$=6.98 sample together with the corresponding quantities from NMR measurements for $^{17}$O and $^{63}$Cu nuclei in YBa$_{2}$Cu$_{3}$O$_{7}$.\cite{O-NMR,Cu-NMR} 
The relaxation rates are plotted for $H\bot c$ orientation and $^{63}$Cu NMR relaxation data is plotted for weak magnetic field $H$=0.45 T where the fluxoid core contribution to relaxation is small.\cite{Cu-NMR} 
The electron and nuclear relaxation rates were normalized to their values above $T_{c}$ at T=100 K. As is evident from Fig.~6, $(1/T_{1}T)$ shows a very similar temperature dependence for the Yb$^{3+}$ electronic spins and the $^{63}$Cu and $^{17}$O nuclei. 

Previously, the relaxation rate $1/T_{1}$ of Yb$^{3+}$ in the superconducting state of optimally doped YBa$_{2}$Cu$_{3}$O$_{7}$ was extracted from $^{170}$Yb M\"ossbauer spectra.\cite{Hodges} The temperature dependence of $1/T_{1}$ was followed only up to 90 K limited by the decrease in the intensity of the M\"ossbauer effect. Therefore, a sharp decrease of the relaxation rate at transition from normal to superconducting state was not observed. Nevertheless, it is interesting to compare the absolute values of the relaxation rates of Yb$^{3+}$ obtained by M\"ossbauer and EPR techniques. Such a comparison is shown in Table II. The excellent quantitative agreement of the relaxation rates obtained by two different experimental techniques is remarkable and provides  strong support of the methods of extracting relaxation rates from M\"ossbauer spectroscopy\cite{Hodges} and from EPR spectra presented in this work.

\begin{table}[htb]

\caption[~]{The relaxation rates of Yb$^{3+}$ in YBa$_{2}$Cu$_{3}$O$_{7}$ at different temperatures obtained by M\"ossbauer technique ($1/T_{1}^{MS}$, Ref.~\onlinecite{Hodges}) and by EPR in the present work ($1/T_{1}^{EPR}$)}  

\begin{center}
\begin{tabular}{ccc}
\hline
\hline
\\ 
$T$(K) & $1/T_{1}^{MS}$(10$^{9}$ s$^{-1}$) & $1/T_{1}^{EPR}$(10$^{9}$ s$^{-1}$)
 \\
\hline
60 & 0.5(1) & 0.6(1)   \\

70 & 1.3(1) & 1.2(1)  \\

80 & 2.5(1) & 1.8(1)   \\

90 & 4.0(1) & 4.4(1)  \\
\hline
\hline
\end{tabular}
\end{center}

\end{table}

Relaxation measurements at low temperatures ($T\ll T_{c}$) can provide information about the superconducting gap symmetry.\cite{Asayama} However, in the present EPR experiments we could not detect relaxation broadening at $T\ll T_{c}$, since it becomes much smaller than the inhomogeneous residual linewidth. This prevents a reliable measurement of the relaxation rate below $T=0.5T_{c}$. It is expected that at low temperatures in the superconducting state the Yb$^{3+}$ relaxation rate will decrease below 10$^{7}$ s$^{-1}$. In this case $T_{1}$ can be measured directly using the pulse EPR techniques. It would be interesting to perform such experiments at low temperatures.

\section{Summary and Conclusions}

To summarize, we performed a detailed study of the temperature dependence of the Yb$^{3+}$ EPR linewidth, i.e., the  relaxation in YBa$_{2}$Cu$_{3}$O$_{x}$ from the undoped insulating to the optimally doped superconducting region ($6\leq x\leq7$). It was found that both electronic and phononic processes contribute to Yb$^{3+}$ relaxation. We were able to separate these processes and studied their relative contributions to relaxation as a function of oxygen doping. As expected, the electronic contribution decreases with decreasing oxygen doping, while the phonon contribution is practically doping independent. It was found that the phononic part of relaxation has an exponential temperature dependence, which cannot be explained by a traditional mechanism involving acoustic phonons. Instead, a Raman process via the coupling to high-energy ($\sim$500 K) optical phonons or an Orbach-like process via the excited vibronic levels of the Cu$^{2+}$ ions (localized Slonczewski-modes) is responsible for the phononic part of the Yb$^{3+}$ relaxation in YBa$_{2}$Cu$_{3}$O$_{6+x}$.

In a sample with  maximum oxygen doping $x=6.98$, the electronic part of relaxation follows the Korringa law in the normal state, and a sharp drop of the relaxation rate was observed below $T_{c}$. Comparison of the EPR linewidths in samples with and without Zn doping allowed us to prove that the superconducting gap opening is responsible for the sharp decrease of Yb$^{3+}$ relaxation in  YBa$_{2}$Cu$_{3}$O$_{6.98}$. It was shown that the electronic part of the Yb$^{3+}$ relaxation rate in the superconducting state follows a very similar temperature dependence as the $^{63}$Cu and the $^{17}$O nuclear relaxation rates, despite the huge difference between the corresponding electronic and nuclear relaxation rates.

There is an excellent quantitative agreement between relaxation rates of Yb$^{3+}$ in YBa$_{2}$Cu$_{3}$O$_{7}$ obtained previously by M\"ossbauer spectroscopy and in the present work by EPR. This provides  strong support of the methods of extracting relaxation rates from M\"ossbauer spectroscopy and from EPR spectra. One should note however, that the EPR signal from Yb$^{3+}$ in YBa$_{2}$Cu$_{3}$O$_{x}$ can be followed up to at least 160 K, while M\"ossbauer measurements in the same compound using $^{170}$Yb are limited by 90 K due to the decrease in the intensity of the M\"ossbauer effect with temperature.

The present results demonstrate that Yb$^{3+}$ can serve as a very effective microscopic spin probe to study electronic, magnetic and lattice properties of YBa$_{2}$Cu$_{3}$O$_{x}$.

\section{Acknowledgments}

This work was supported by the Swiss National Science Foundation, the SCOPES grant No. IB7420-110784, and the NCCR program MaNEP.


\begin{thebibliography}{99}

%
\bibitem{Hor} P.H. Hor, R.L. Meng, Y.Q. Wang, L. Gao, Z.J. Huang, J. Bechtold, K. Forster, and C.W. Chu, Phys. Rev. Lett. {\bf58}, 1891 (1987).

\bibitem{Mesot} J. Mesot and A. Furrer, J. Supercond. {\bf10}, 623 (1997).

\bibitem{Boothroyd1} A.T. Boothroyd, Phys. Rev. B {\bf64}, 066501 (2001).

\bibitem{Staub1} S.W. Lovesey and U. Staub, Phys. Rev. B {\bf64}, 066502 (2001).

\bibitem{Walter} U. Walter, S. Fahy, A. Zettl, S.G. Louie, M.L. Cohen, P. Tejedor, and A.M. Stacy, Phys. Rev. B {\bf36}, 8899 (1987).

\bibitem{Amoretti} G. Amoretti, R. Caciuffo, P. Santini, O. Francescangeli, E.A. Goremychkin, R. Osborn, G. Calestani, M. Sparpaglione, and L. Bonoldi, Physica C {\bf221}, 227 (1994).

\bibitem{Boothroyd2} A. Mukherjee, A.T. Boothroyd, D. MK. Paul, M. P. Sridhar Kumar, and M. A. Adams, Phys. Rev. B {\bf49}, 13089 (1994); A.T. Boothroyd, A. Mukherjee, and A.P. Murani, Phys. Rev. Lett. {\bf77}, 1600 (1996).

\bibitem{Furrer} J. Mesot, G. B\"ottger, H. Mutka, and A. Furrer, Europhys. Lett. {\bf44}, 498 (1998).

\bibitem{Staub2}  U. Staub, M. Gutmann, F. Fauth, and W. Kagunya, J. Phys.: Condens. Matter {\bf11}, L59 (1999).

\bibitem{Staub3}  S.W. Lovesey and U. Staub, Phys. Rev. B {\bf61}, 9130 (2000).

\bibitem{Roepke}  M. Roepke, E. Holland-Moritz, B. B\"uchner, H. Berg, R.E. Lechner, S. Longeville, J. Fitter, R. Kahn, G. Coddens, and M. Ferrand, Phys. Rev. B {\bf60}, 9793 (1999).

\bibitem{Barnes} S.E. Barnes, Adv. Phys. {\bf30}, 801 (1981).

\bibitem{Elschner}  L. Kan, S. Elschner, and B. Elschner, Solid State Commun. {\bf79}, 61 (1991).

\bibitem{Eremin} M.V. Eremin, I.N. Kurkin, M.P. Rodionova, I.H. Salikhov, and L.R. Tagirov, Supercond. Phys. Chem. Techn. (Russia) {\bf4}, 716 (1991).

\bibitem{Kurkin}  I.N. Kurkin, I.Kh. Salikhov, L.L. Sedov, M.A. Teplov, and R.Sh. Zhdanov, JETP {\bf76}, 657 (1993).

\bibitem{Shimizu1}  H. Shimizu, K. Fujiwara, and K. Hatada, Physica C {\bf288}, 190 (1997).

\bibitem{Shimizu2}  H. Shimizu, K. Fujiwara, and K. Hatada, Physica C {\bf299}, 169 (1998).

\bibitem{Ivanshin1}  V.A. Ivanshin, M.R. Gafurov, I.N. Kurkin, S.P. Kurzin, A. Shengelaya, H. Keller, and M. Gutmann, Physica C {\bf307}, 61 (1998).

\bibitem{Ivanshin2}  L.K. Aminov, V.A. Ivanshin, I.N. Kurkin, M.R. Gafurov, I.Kh. Salikhov, H. Keller, and M. Gutmann, Physica C {\bf349}, 30 (2001).

\bibitem{Gafurov}  M.R. Gafurov, L.K. Aminov, I.N. Kurkin, and V.V. Izotov, Supercond. Sci. Technol. {\bf18}, 352 (2005).

\bibitem{Farrell} D.E. Farrell, B.S. Chandrasekhar, M.R. DeGuire, M.M. Fang, V.G. Kogan, J.R. Clem, and D.K. Finnemore, Phys. Rev. B {\bf36}, 4025 (1987).

\bibitem{Bleaney} A. Abragam and B. Bleaney, Electron Paramagnetic
Resonance of Transition Ions (Clarendon Press, Oxford, 1970).

\bibitem{Guillaume} M. Guillaume, P. Allenspach, J. Mesot, U. Staub, A. Furrer, R. Osborn, A.D. Taylor, F. Stucki, and P. Untern\"ahrer, Solid State Commun. {\bf81}, 999 (1992).

\bibitem{Kochelaev} J. Sichelschmidt, B. Elschner, A. Loidl, and B.I. Kochelaev,  Phys. Rev. B {\bf51}, 9199 (1995).

\bibitem{Smirnov} A. I. Smirnov and R. L. Belford, J. Magn. Reson. A {\bf113}, 65
(1995).

\bibitem{Stapleton} Gh. Cristea, T. L. Bohan, and H. J. Stapleton, Phys. Rev. B {\bf4}, 2081 (1971).

\bibitem{Atsarkin1} V.A. Atsarkin, V.V. Demidov, G.A. Vasneva, T. Feher, A. J\'anossy, and B. Dabrowski, Phys. Rev. B {\bf61}, R14944 (2000).


\bibitem{Orbach} S.A. Dodds, J. Sanny, and R. Orbach, Phys. Rev. B {\bf18}, 1016 (1978).

\bibitem{Kochelaev1} B.I. Kochelaev, JETP {\bf10}, 171 (1960).

\bibitem{Huang} Chao-Yuan Huang, Phys. Rev. {\bf154}, 215 (1967).

\bibitem{Pintschovius} L. Pintschovius, W. Reichardt, M. Klaser, T. Wolf, and H.V. Lohneysen, Phys. Rev. Lett. {\bf89}, 037001 (2002); M. Opel, R. Hackl, T.P. Devereaux, A. Virosztek, A. Zawadowski, A. Erb, E. Walker, H. Berger, and L. Forro, Phys. Rev. B {\bf60}, 9836 (1999).


\bibitem{Muller1} U. H\"ochli and K.A.~M\"uller, Phys. Rev. Lett. {\bf12}, 730 (1964).

\bibitem{Muller2} U. H\"ochli, K.A.~M\"uller, and P. Wysling, Phys. Lett. {\bf15}, 1 (1965).

\bibitem{Muller3} K.A.~M\"uller, in {\it Magnetic Resonance and Relaxation}, edited by R. Blinc (North-Holland Publishing Company, 1967), pp. 192-208.

\bibitem{Alloul} H. Alloul, T. Ohno, and P. Mendels, Phys. Rev. Lett. {\bf63}, 1700 (1989).

\bibitem{Janossy} A. J\'anossy, L.C. Brunel, and J.R. Cooper, Phys. Rev. B {\bf54}, 10186 (1996).

\bibitem{Shaltiel} D. Shaltiel, C. Noble, J. Pilbrow, D. Hutton, and E. Walker, Phys. Rev. B {\bf53}, 12430 (1996).

\bibitem{Garifullin} N.E. Alekseevskii, A.V. Mitin, V.I. Nizhankovskii, I.A. Garifullin, N.N. Garifyanov, G.G. Khaliullin, E.P. Khlybov, B.I. Kochelaev, and L.R. Tagirov, J. Low Temp. Phys. {\bf77}, 87 (1989).

\bibitem{Kochelaev-Review} B.I. Kochelaev and G.B. Teitelbaum, in {\it Superconductivity in Complex Systems}, (Springer, 2005), pp. 205-266.

\bibitem{Atsarkin} V.A. Atsarkin, V.V. Demidov, and G.A. Vasneva, Phys. Rev. B {\bf52}, 1290 (1995).

\bibitem{O-NMR} J. A. Martindale, P.C. Hammel, W. L. Hults, and J. L. Smith, Phys. Rev. B {\bf57}, 11769 (1998).

\bibitem{Cu-NMR} J. A. Martindale, S. E. Barrett, C. A. Klug, K. E. O'Hara, S. M. DeSoto, C. P. Slichter, T. A. Friedman, and D. M. Ginsberg, Phys. Rev. Lett. {\bf68}, 702 (1992).

\bibitem{Hodges} J. A. Hodges, P. Bonville, P. Imbert, and G. J\'ehanno,  Physica C {\bf184}, 259 (1991).

\bibitem{Asayama} For a review see, K. Asayama, Y. Kitaoka, G.Q. Zheng, and K.
Ishida, Prog. Nucl. Magn. Reson. Spectrosc. {\bf87}, 221 (1996).

\end{thebibliography}
\end{document}